\documentclass{aa}
\usepackage{natbib}
\bibpunct{(}{)}{;}{a}{}{,}
\usepackage[varg]{txfonts}
\usepackage{graphicx}
\usepackage{hyperref}
\usepackage[normalem]{ulem}

\newcommand{\msun}{\mbox{M$_{\odot}$}}

\newcommand{\degs}{\ifmmode ^{\circ}\else$^{\circ}$\fi}
\newcommand{\amin}{\ifmmode ^{\prime}\else$^{\prime}$\fi}
\newcommand{\asec}{\ifmmode ^{\prime\prime}\else$^{\prime\prime}$\fi}

\title{There and back again: Mysterious optical pulse profile behavior of the transitional millisecond pulsar PSR J1023+0038}
\author{S. V. Karpov\inst{1}
\and Artyom S. Tanashkin\inst{2}\fnmsep\thanks{E-mail: artyom.tanashkin@gmail.com}
\and G. M. Beskin\inst{3}
\and V. L. Plokhotnichenko\inst{3}
\and \\Y. A. Shibanov\inst{2}
\and D. A. Zyuzin\inst{2}
}

\institute{Institute of Physics of the Czech Academy of Sciences, CZ-182 21 Prague 8, Czech Republic
\and Ioffe Institute, Politekhnicheskaya 26, St. Petersburg, 194021, Russia
\and Special Astrophysical Observatory, Nizhny Arkhyz, 369167, Russia
}

\date{Received date /
Accepted date }

\abstract{Neutron stars in close binary systems have the potential to spin up to millisecond periods due to the accretion of matter and angular momentum from their low-mass companions. In later stages of this process, they sometimes start to swing between the accretion-powered and rotation-powered regimes, manifesting themselves as low-mass X-ray binaries and millisecond radio pulsars, respectively. Such systems are known as transitional millisecond pulsars. PSR J1023+0038 was the first one of this kind to be discovered and the first to show optical pulsations at the rotation frequency of the neutron star during a peculiar low accretion state. The optical pulse profile is characterized by a smooth double-peaked shape resembling thermal light curves of X-ray pulsars, but more likely emerging from re-emission of the pulsar wind energy by charged particles in the surrounding medium. Although the amplitudes of the peaks show strong variability, the overall structure of the pulse profile was observed to be fixed, with the optical pulsed fraction $\lesssim 1$ percent. Here, we report time-resolved observation of a dramatic violation of this permanence during our high temporal resolution observations of PSR J1023+0038 with the 6-m BTA telescope of the Special Astrophysical Observatory. On a timescale of a few seconds the pulse profile took a single-peaked, nearly sinusoidal shape, with synchronous magnification of the pulsed fraction up to $\approx 5$ percent. After spending $\approx 220$ seconds in this new regime, accompanied by flaring activity, the system returned to its normal state. We discuss a number of possible explanations of this peculiar behavior in the context of the physics and geometry of interactions between the pulsar wind and surrounding matter. However, a complete picture is yet to be revealed.}


\titlerunning{PSR J1023+0038 pulse profile transformation}
\authorrunning{S. V. Karpov et al.}
\begin{document}

\maketitle
\nolinenumbers

\section{Introduction}
\label{sec:intro}

Millisecond pulsars (MSPs) constitute a subclass of rapidly rotating neutron stars (NSs) that demonstrate coherent millisecond pulsations in one or more energy ranges from the radio to gamma-rays. The first representative of this class, radio pulsar PSR B1937+21, was discovered in 1982 \citep{Backer+1982} and its nature was immediately explained on the basis of the recycling scenario \citep{Alpar+1982,Radhakrishnan+1982}. According to this view, first introduced theoretically by \citet{BKK1974}, the NS is spun-up by accretion of matter and angular momentum from a low-mass companion star in a close binary system. During the accretion stage, the NS is expected to exhibit X-ray pulsations due to magnetic channeling of the accreting matter and release of its kinetic energy in the magnetic pole regions. Thus, the discovery of a recycled accreting X-ray MSP, SAX J1808.4-3658 \citep{Wijnands+1998}, in 1998 was a strong confirmation of the presented picture \citep[see also detection of millisecond pulsations from 4U 1728-34 during X-ray bursts by][]{Strohmayer+1996}.
        
However, the most compelling evidence came with the discovery of transitional MSPs. The first one, \object{PSR J1023+0038} (hereafter J1023), was initially classified in early 2000s as a cataclysmic variable due to its radio emission, optical variability and prominent double-peaked emission lines indicating presence of an accretion disk \citep{Bond+2002,Szkody+2003}. Since early 2002, the source has shown no further sign of accretion or emission lines \citep[e.g.,][]{Thorstensen+2005,Homer+2006}. Moreover, in June 2007 a bright MSP was found in the radio \citep{Archibald+2009}, undoubtedly indicating that the system performed the transition from the low-mass X-ray binary (LMXB) to the radio pulsar state. For six years, J1023 behaved like a typical redback pulsar \citep{Archibald+2010,Bogdanov+2011,Archibald+2013}, but in June 2013 it transitioned back to the subluminous disk state \citep{Patruno+2014,Stappers+2014} where it has remained since then. At the moment, only two other transitional MSPs have been conclusively identified, IGR J18245-2452 \citep{Papitto+2013} and XSS J12270-4859 \citep{Bassa+2014,Roy+2015}. A few candidates with a similar morphology await confirmation \citep{Bogdanov&Halpern2015,Strader+2016,Li+2020,Coti+2019,Miller+2020,Strader+2021,Kyer+2025,Zyuzin+2025}.

\begin{table*}
\renewcommand{\arraystretch}{1.2}
\caption{Summary of observations of J1023 with the BTA telescope.}
\label{tab:obs}
\begin{center}
\begin{tabular}{ccccc}
\hline
Night & Date (dd.mm.yyyy) & Date (MJD) & Start time (hh:mm:ss, UT) & Exposure (s) \\
\hline
1 & 14.11.2017 & 58071 & 02:00:21 & 3200 \\
2 & 15.11.2017 & 58072 & 00:54:19 & 6700 \\
\hline
\end{tabular}
\end{center}
\end{table*}

J1023 is characterized by extremely rich variety of observational properties revealed  through a numerous multi-wavelength campaigns in recent years. We refer to Table~2 from \citet{Jaodand+2016} and Table~6.1 from \citet{Papitto_deMartino_2022} for a summary of its main parameters. In the LMXB state, J1023 exhibits sporadic switches between three different X-ray modes \citep{Bogdanov+2015}. About 70 percent of the time, it remains in the high mode with the X-ray luminosity in the 0.3--10~keV band $L_\mathrm{X} \simeq 3\times10^{33}$~erg s$^{-1}$, with rapid switches to the low mode ($L_\mathrm{X}\simeq5\times10^{32}$~erg s$^{-1}$) and occasional flares ($L_\mathrm{X}\simeq10^{34}$~erg s$^{-1}$), also observed in the optical band \citep{Papitto+2019}. These luminosities are considerably lower than the values typical for accreting MSPs during outbursts \citep[$\sim 10^{36}-10^{37}$~erg s$^{-1}$, ][]{Campana+2018}. Moreover, for the NS mass of $1.7$~\msun\, \citep{Deller+2012} and typical radius of 10 km, the corresponding accretion rates $\dot{\mathfrak{M}} \simeq 2 \times 10^{12} - 4 \times 10^{13} \text{g s}^{-1}$ are insufficient to overcome the centrifugal barrier imposed by a rapidly rotating magnetosphere of the NS \citep{Archibald+2015,Papitto+2019}. Coherent X-ray pulsations are detected only during the high mode \citep{Archibald+2015,Papitto+2019}, although \citet{Bogdanov+2015} reported a marginal (with a single-trial significance of $\sim2\sigma$) signal in the flare mode. 

The isotropic optical luminosity in the 320--900~nm band is $L_\mathrm{opt}\sim10^{33}$~erg s$^{-1}$ \citep{Papitto+2019}. In the optical and infrared, J1023 displays sporadic activity on seconds to hours time scales \citep{Shahbaz+2015,Shahbaz+2018,Hakala+2018,Kennedy+2018,Messa+2024}. The minimal stochastic variability time scale of the optical emission is as short as fractions of seconds \citep{Shibanov+2017,Beskin+2018},  which is close to the characteristic time scales of variability due to matter fragmentation in a propeller regime demonstrated by MHD simulations \citep{Romanova+2018}. The presence of some kind of an outflow is suggested by highly variable, flat-spectrum continuum radio emission anti-correlated with  X-rays \citep{Deller+2015,Bogdanov+2015,Bogdanov+2018}. 

Surprisingly, coherent optical \citep{Ambrosino+2017,Zampieri+2019,Karpov+2019} and near-UV \citep{Jaodand+2021,Miraval+2022} pulsations at the NS spin frequency were detected. Simultaneous X-ray and optical observations \citep{Papitto+2019} indicated that optical pulsations are present during the high X-ray mode, having a double-peaked profile similar to that seen in X-rays. Analogous behavior is observed in the UV \citep{Miraval+2022}. Pulsations in all three energy ranges are likely produced in regions located within a few kilometers from each other and by the same mechanism. \citet{Papitto+2019} also detected optical pulsations with reduced amplitude in the flare mode.

A number of models have been proposed to explain such a wealth of observational phenomena seen in J1023. The main questions include whether the ejector mechanism continues to operate in the presence of the accretion disk, how strong the propeller mechanism is, whether a significant fraction of matter ends up reaching the NS surface, what the origins of the X-ray and optical emission are, and what process governs the switches between high, low, and flaring X-ray modes \citep[see e.g.][]{Coti+2014,Li+2014,Stappers+2014,Takata+2014,Papitto+2015,Campana+2016}. The discovery of the optical pulsations essentially ruled out accretion onto the NS surface as the mechanism behind the generation of the pulsing emission in the optical and X-rays. \citet{Papitto+2019} and \citet{Veledina+2019} suggested instead that such emission originates in the shock powered by collision between the pulsar wind and the disk just beyond the light cylinder during the high mode. A number of observational works generally support this scenario \citep[][also see however \citealt{Jadoliya+2026}]{Campana+2019,Burtovoi+2020,Illiano+2023,Baglio+2023,Baglio+2025apj,Baglio+2025aap}. We discuss this in more detail in Sect.~\ref{subsec:origin}.

In 2017--2020, we carried out an observational campaign with the 6-m BTA telescope, focusing on the coherent optical millisecond pulsations of J1023. The system was observed over seven nights, providing about 10 hours of total exposure time. Here, we present an analysis of the data obtained over two consecutive nights on November 14--15, 2017, featuring a puzzling behavior of the optical pulse profile never seen before in pulsars with smooth, isotropic-like light curves. The rest of the data, along with the general properties of observed millisecond pulsations and subsecond flaring activity will be discussed in subsequent papers. 

Observations and data reduction are described in Sect.~\ref{sec:data}, while the results are presented in Sect.~\ref{sec:results}. In Sect.~\ref{sec:discussion}, we discuss possible scenarios capable of explaining the observed behavior of the system. We draw our conclusions in Sect.~\ref{sec:sum}. In Appendix~\ref{a:swift}, we briefly consider \textit{Swift/XRT} X-ray observations of J1023 two weeks after our data set.

\section{Observations and data reduction}
\label{sec:data}

We observed J1023 with the Multi-channel panoramic photo-polarimeter \citep[MPPP, ][]{Plokhotnichenko+2021} mounted on the 6-m BTA telescope of the Special Astrophysical Observatory of the Russian Academy of Sciences (SAO RAS). The observation log is presented in Table~\ref{tab:obs}. The device operated in dual-channel regime allowing to collect photons simultaneously with the two MCP-based panoramic photon counters, the ``red'' one with the GaAs photocathode sensitive at the 5670 {\AA} effective wavelength, and the ``blue'' one with the multi-alkali photocathode sensitive at the 4320 {\AA} effective wavelength. Photons in a $10\asec \times 10\asec$ diaphragm around the object were detected and registered with the effective temporal resolution of $\sim 1\,\mu$s.

For to technical reasons, no comparison star was observed, precluding the standard flux correction procedure. To roughly estimate the total background contribution, which is essential for calculation of the pulsed fraction of the source, we considered 10 second-long frames, taking the mean flux in the source-free area of each frame as the sky contribution to the source flux, similarly to what is usually done in X-ray astronomy. We note that this approach is possible since we are only interested in the total background contribution to the pulse profile, rather than its short-timescale fluctuations. As a result, we estimated the number of the background photons as $3461883\pm1860$ (0.47 of the total) and $945797\pm973$ (0.34 of the total) in the red and blue channels for the first night, and $6074874\pm2465$ (0.38 of the total) and $1398271\pm1182$ (0.24 of the total) in the red and blue channels for the second night.

We followed the standard procedure to search for periodic signal in our data. The photon times of arrival (ToAs) were converted to the barycenter of the Solar system using DE405 ephemeris. To account for orbital motion of the pulsar in the binary system, the ToAs are usually converted to the epoch of passage at the ascending node, $T_\mathrm{asc}$, 
\begin{equation}\label{eq:tcorr}
    t = t_0 - a_0\sin 2\pi\varphi,
\end{equation}
where $t_0$ and $t$ are the ToAs before and after correction, $\varphi = 2\pi(t-T_\mathrm{asc})/P_\mathrm{orb}$ is the orbital phase of the NS\footnote{For practical calculations, in this formula, $t$ is usually replaced by $t_0$, with the resulting error on the NS position on the order of the distance spanned by the NS during the time required for light to travel across the orbit.}, $a_0$ is the projection of the semi-major axis of the orbit to the line of sight\footnote{In the case of J1023, the orbit is approximately circular and $a_0$ can be approximated as a projection of the orbital radius to the line of sight.}, and $P_\mathrm{orb}$ is the orbital period.

\begin{figure}
\begin{minipage}[h]{1.\linewidth}
\center{\includegraphics[width=1.0\linewidth,trim={0 0 0 2cm},clip]{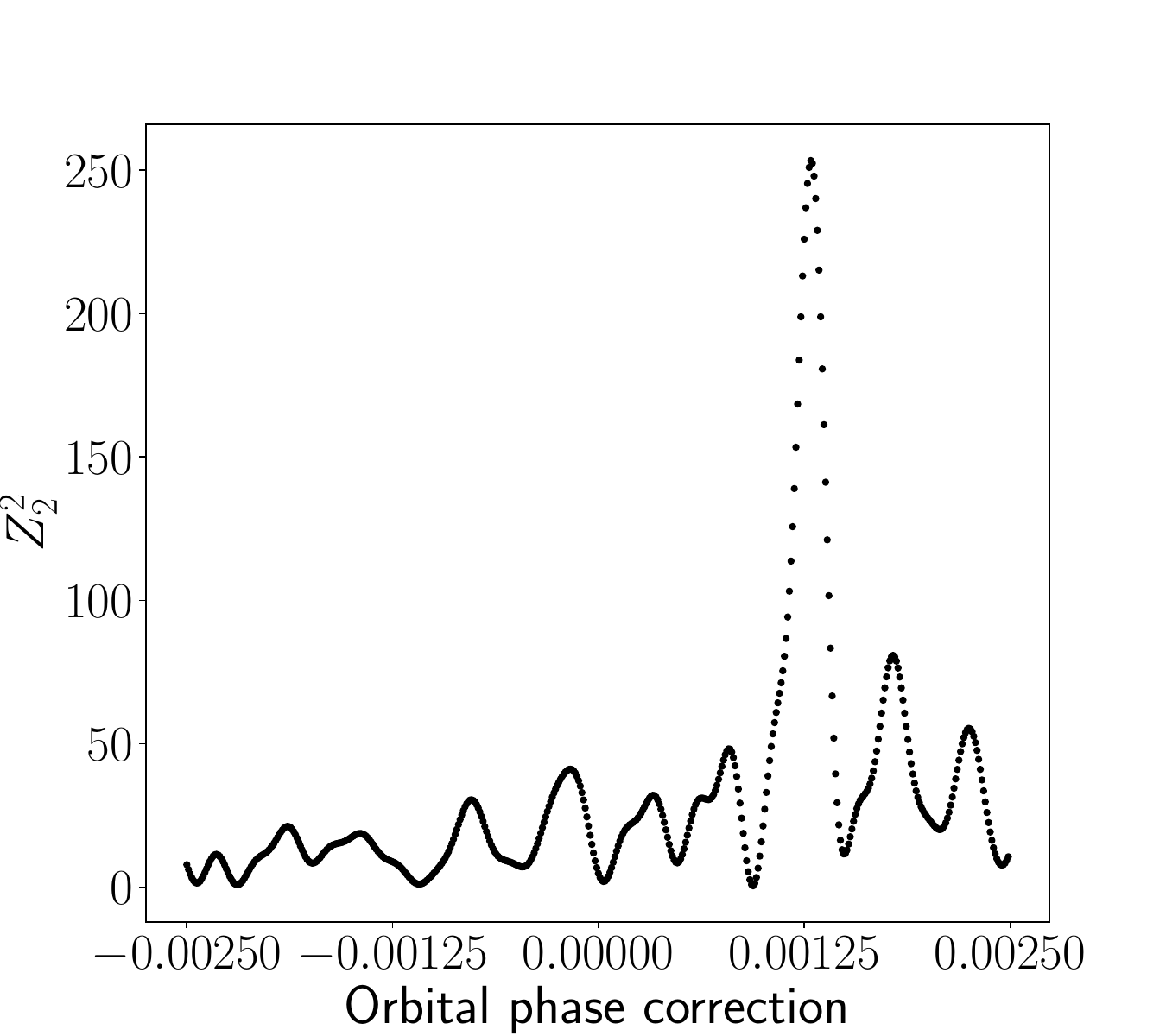}}
\end{minipage}
\caption{$Z_2^2$-test for both nights joined. The orbital phase correction is shown with respect to the ephemeris value obtained from the timing solution by \citet{Jaodand+2016}. The maximum at 0.00129 corresponds to the shift of the epoch of passage at the ascending node by $-22$ seconds with respect to 54905.97140075 MJD. The aliases around the main peak emerge due to the presence of gaps in each data set and between them.}
\label{fig:ztest}
\end{figure}

The orbital phase of J1023 (or, equivalently, the time of passage at the ascending node) is known to exhibit rather peculiar behavior. After the transition in 2014 from the radio pulsar state to the subluminous disk state, the time of passage at the ascending node shifted by about 25 seconds (back in time) and then for a few years varied around this value \citep[e.g.][]{Jaodand+2016,Zampieri+2019,Papitto+2019}. Sometime in 2019 these variations turned into an increasing trend, as was shown by \citet{Illiano+2023} using \textit{NICER} observations and recently confirmed by \citet{Conforti+2026} on a larger data set (see their Fig.~1), which could be associated with underestimation of the orbital period. In any case, fine tuning of the orbital phase is required. We calculated the ephemeris orbital phase based on the solution by \citep{Jaodand+2016} and considered a set of corrections in the interval from $-0.0025$ to $0.0025$ with the resolution of $10^{-5}$. This interval covers all stochastic variations of the orbital phase ever observed in J1023. For each correction we calculated the $Z_K^2$-statistic \citep{Buccheri+1983} as
\begin{equation}\label{eq:ztest}
    Z_K^2 = \frac{2}{N}\sum_{k=1}^{K}\left[\left(\sum_{i=1}^{N}\cos 2\pi k\phi_i\right)^2+\left(\sum_{i=1}^{N}\sin 2\pi k\phi_i\right)^2\right],
\end{equation}
taking into account the first two harmonics ($K=2$) of the pulsar frequency since the optical pulse profile is known to have a smooth double-peaked structure. Here, $N$ is the total number of collected photons and $\phi_i$ is the rotation phase of the NS corresponding to the ToA of the $i^\mathrm{th}$ photon, which depends on the assumed pulsar frequency and orbital phase (through the ToAs correction given by Eq.~\ref{eq:tcorr}). The orbital period of the system and the projection of the semi-major axis of the orbit to the line of sight were fixed at the values reported in \citet{Jaodand+2016}. The pulsar frequency was fixed at $f = 592.421467517559$~Hz, which is the \citet{Jaodand+2016} frequency corrected for the epoch difference using the frequency derivative $\dot{f} = -3.0413\times10^{-15}$~Hz s$^{-1}$. The result is shown in Fig.~\ref{fig:ztest}. A strong periodic signal ($Z_2^2\simeq250$) is found at the orbital phase correction of 0.00129(9), translating to $-22$ s shift of the time of passage at the ascending node with respect to 54905.97140075 MJD \citep{Jaodand+2016}. The associated uncertainty is the half width at half maximum (HWHM) of a Gaussian fitted to the peak. The corrected value at our epoch (58072 MJD) is $T_\mathrm{asc} = 54905.97115(2)$~MJD. The observations were taken in 2017, so this value is compatible with the behavior observed back then. The significance of the peak is very high. For example, for a $5\sigma$ confidence level with 500 independent trials, the detection threshold for $Z_2^2$ (which for the Poisson noise is distributed as $\chi^2$ with 4 degrees of freedom) is 48. We also checked that the ephemeris pulsar frequency is indeed the optimal frequency in terms of the signal strength and no correction is needed in this regard.

After the specified reduction and optimization of orbital parameters, we constructed binned pulse profiles of the source at the rotation period. The numbers of phase bins were estimated using Eq.~1 from \citet{Becker+1999}, which compromises between the pulse profile resolution and signal to noise ratio. 

\begin{figure}
\begin{minipage}[h]{1.\linewidth}
\center{\includegraphics[width=1.0\linewidth,clip]{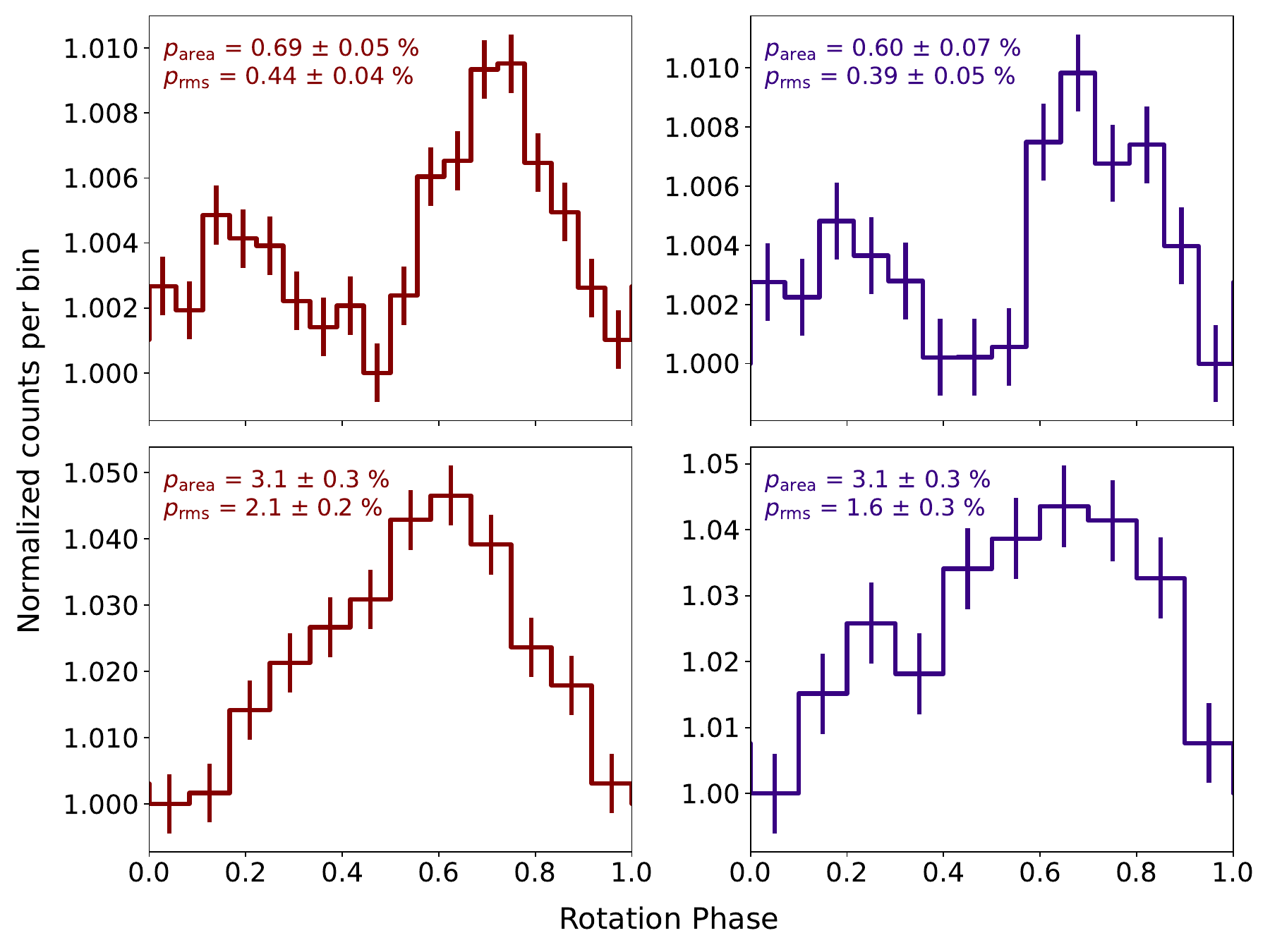}}
\end{minipage}
\caption{Top: Typical double-peaked pulse profiles obtained by phase folding together the Nov 14 and Nov 15 light curves and excluding the single-peaked interval. The red and blue channels are shown on the left and right, respectively. Bottom: Averaged single-peaked pulse profiles of the 220-second-long interval. Corresponding pulsed fractions $p_\mathrm{area}$ and $p_\mathrm{rms}$ are shown in each panel.}
\label{fig:profiles}
\end{figure}

Preliminary search for periodic signal in the data with a 100-second-wide sliding window performed by \citet{Karpov+2019} revealed one short interval of peculiar activity with strong, single-peaked pulsations. Having that in mind, in addition to folding the entire data set, we also analyzed individual chunks of the data on time scales from tens to hundreds of seconds.

To characterize the strength of the pulsing component, we calculated the background-corrected pulsed fractions in the obtained pulse profiles. Different quantities are used in the literature as a representation of the pulsed fraction \citep[see e.g. Appendix C of ][for discussion]{Hare+2021}. We define $p_\mathrm{area}$ as the ratio of the area between the pulse profile and the lowest phase bin level to the full area under the curve,
\begin{equation}
    p_\mathrm{area} = \frac{C-c}{C-b}, 
\end{equation}
where $C$ is the total amount of counts, $c$ is the unpulsed level equal to the number of counts in the lowest phase bin times the number of bins, and $b$ is the estimated total amount of background counts. For comparison with previous studies, we also provide the values of the root mean square (rms) pulsed fraction, $p_\mathrm{rms}$,
\begin{equation}
    p_\mathrm{rms} = \left(\frac{Z_2^2}{N}\right)^{1/2},
\end{equation}
which essentially characterizes the rms amplitude of the harmonic representation of the pulse profile \citep[see also e.g. ][for a similar definition]{Papitto+2019}. We note that $p_\mathrm{rms}$ is systematically lower than $p_\mathrm{area}$ \citep[by a factor of $\sqrt{2}$ for a sinusoidal signal; ][]{Hare+2021}.

\section{Results}
\label{sec:results} 
 
\begin{figure}
\begin{minipage}[h]{1.\linewidth}
\center{\includegraphics[width=1.0\linewidth,clip]{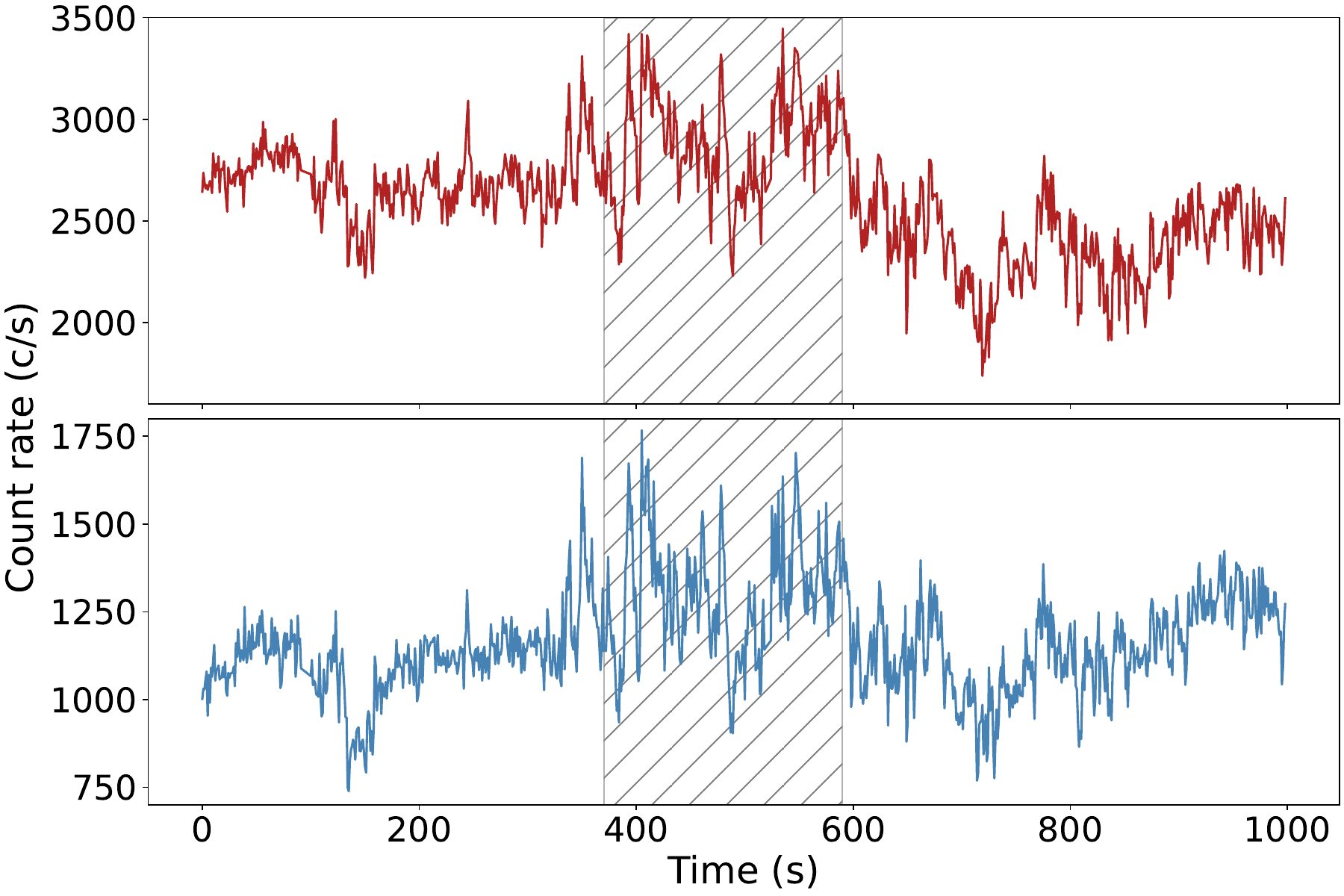}}
\end{minipage}
\caption{Segment of the light curve in the red (top) and blue (bottom) channels containing the interval with single-peaked pulsations (shown with a hatched stripe). Flaring activity around the interval is readily seen. The time resolution is 1 s. Zero time corresponds to 15 Nov 2017 02:37:29 UT.}
\label{fig:flaring}
\end{figure} 

For most of our data, we observed a typical double-peaked pulse profile very similar to those reported in a number of previous studies \citep{Ambrosino+2017,Zampieri+2019,Papitto+2019,Burtovoi+2020,Illiano+2023}. The averaged background-corrected pulsed fractions for the red and blue channels combined (all the detected photons considered together) are $p_\mathrm{area} = 0.65 \pm 0.04$ percent and $p_\mathrm{rms} = 0.43 \pm 0.03$ percent. The two peaks are approximately half of a phase apart, with the first one about two times higher than the second, although the heights of the peaks show strong variability on shorter timescales. In contrast to previous studies, the system was observed simultaneously in two optical bands. The ``normal'' pulse profiles in the red and blue channels are shown on the top panels of Fig.~\ref{fig:profiles}.

\begin{figure}
\begin{minipage}[h]{1.\linewidth}
\center{\includegraphics[width=1.0\linewidth,clip]{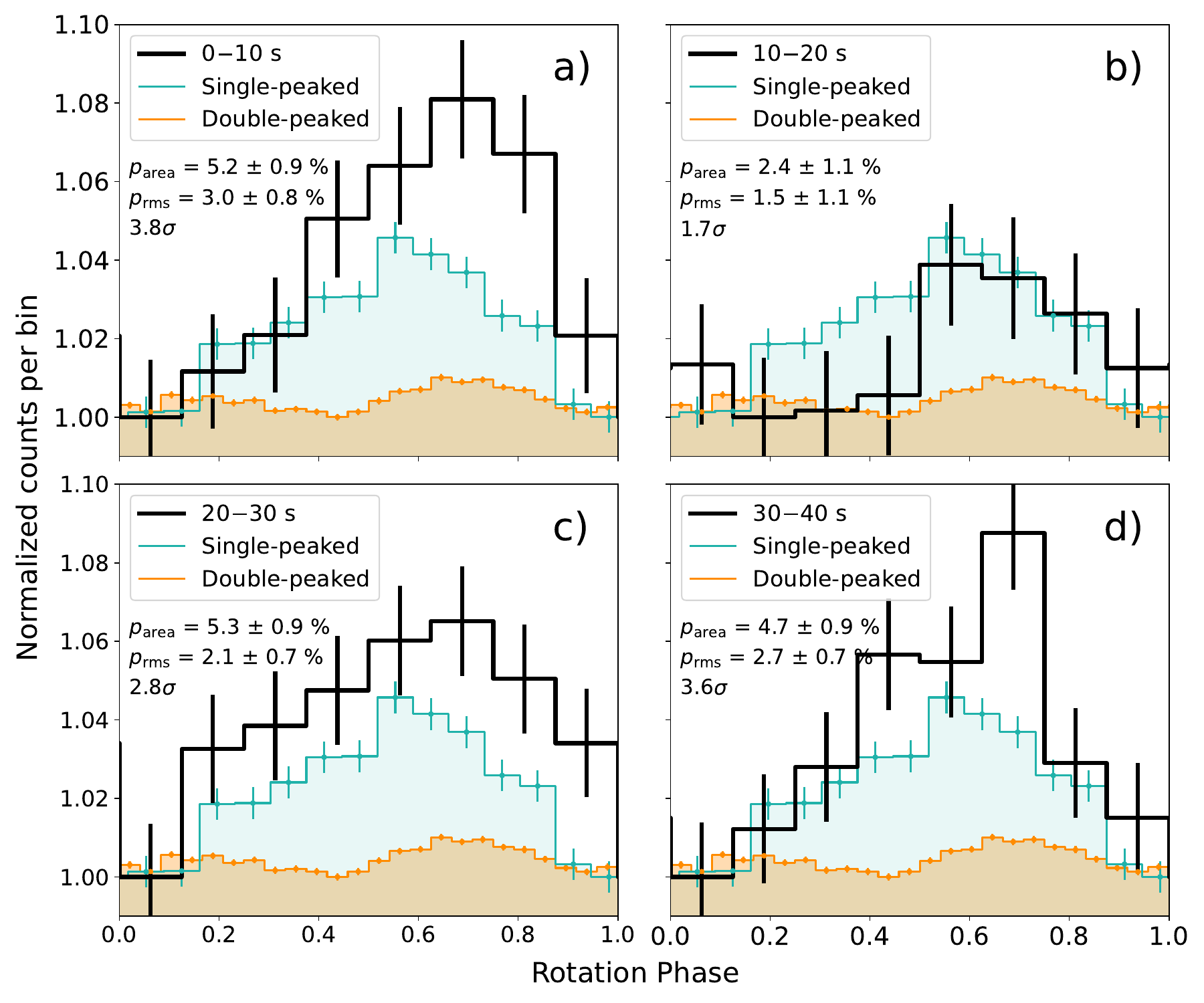}}
\end{minipage}
\caption{Pulse profile variability during the first four 10-second-long segments (a, b, c, and d) of the single-peaked interval. Profiles observed within the time ranges indicated in the legends are plotted in black. Time is counted from the onset of the single peak. Pulsed fractions and detection significance levels are shown in each panel. Blue and orange filled step plots show the averaged pulse profile of the full single-peaked interval and the averaged over both nights typical double-peaked pulse profile, respectively.}
\label{fig:iXframes}
\end{figure}

However, toward the end of the second night an outstanding 220-second-long event took place, starting at 58072.11364~MJD. In the bottom panels of Fig.~\ref{fig:profiles}, the pulse profiles of this interval are shown. The first thing that catches an eye is a completely different pulse shape with a broad, smooth, slightly skewed single peak. It is important to note that the maximum of this peak does not match either of the two in the normal state. The averaged background-corrected pulsed fractions for the sum of the red and blue channels are $p_\mathrm{area} = 3.3 \pm 0.2$ percent and $p_\mathrm{rms} = 1.9 \pm 0.1$ percent (i.e., about four to five times higher than that of the double-peaked pulse profile). Inspecting the light curve of the source, we found that the single-peaked interval is encompassed by flaring activity (Fig.~\ref{fig:flaring}). While flares are consistently observed in J1023 \citep[e.g.][]{Shibanov+2017,Beskin+2018,Shahbaz+2018,Papitto+2019}, we note that neither the two nights under consideration nor any other data we obtained in 2017--2020 contain similar single-peaked and/or high-amplitude pulsation intervals.

\begin{figure}
\begin{minipage}[h]{1.\linewidth}
\center{\includegraphics[width=1.0\linewidth,trim={0 0 0 2cm},clip]{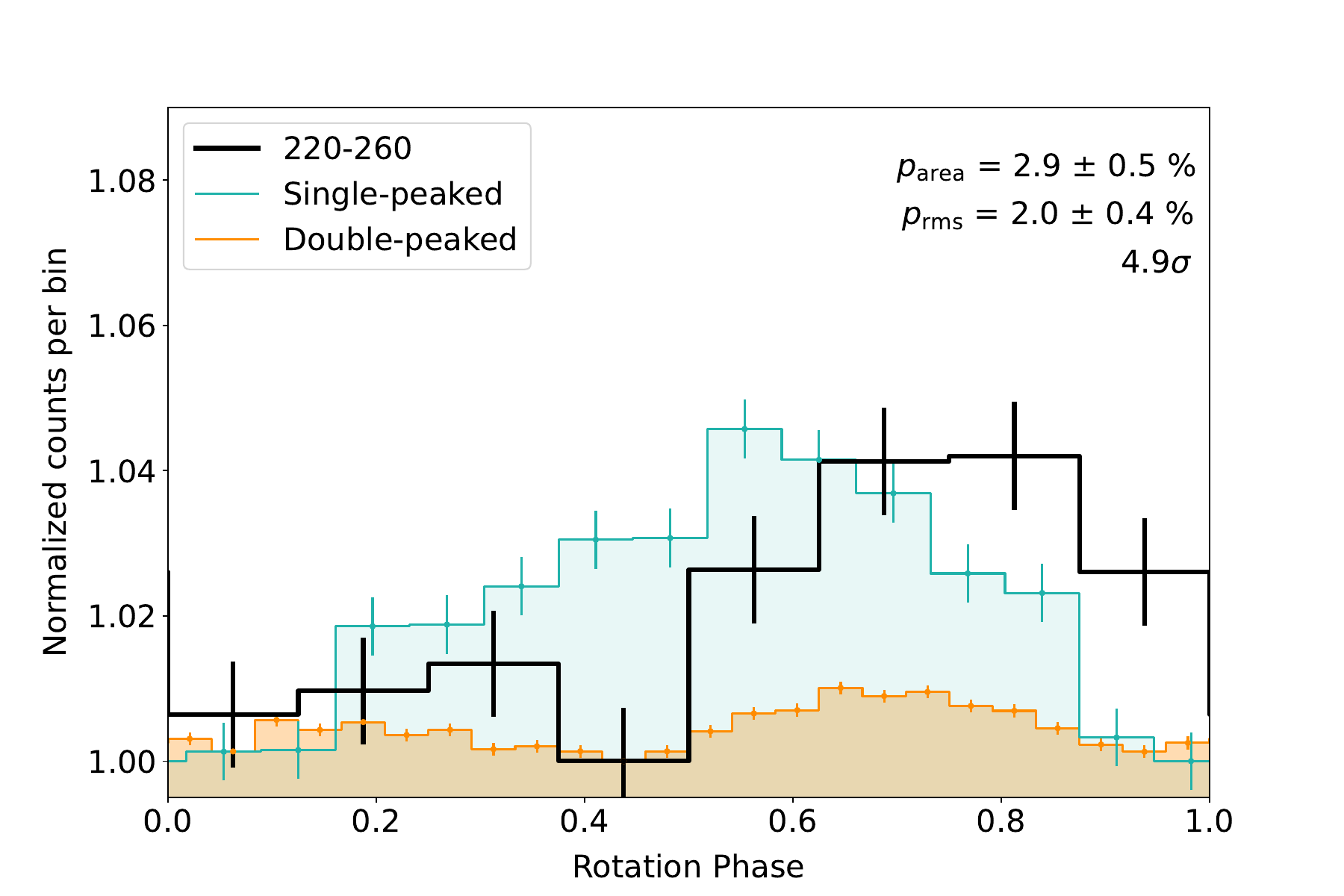}}
\end{minipage}
\caption{Same as Fig.~\ref{fig:iXframes}, but for the first 40 seconds after the single-peaked interval. The pulsed fraction remains high, but the peak is shifted in phase and the pulse profile starts to resemble the normal double-peaked one, although the lower peak is lost in noise at such a short integration time.}
\label{fig:afteriX}
\end{figure}

Thanks to much stronger signal, we are able to investigate temporal variability of the pulse profile during this event at timescales as short as about 10 seconds. Figure~\ref{fig:iXframes} provides an example of such variability. Prominent pulsations seen in the first 10 seconds of the event (panel a of Fig.~\ref{fig:iXframes}, $p_\mathrm{area}=5.2\pm0.9$ percent, with a detection significance of $3.8\sigma$) almost cease in noise in the next 10 seconds (panel b, $p_\mathrm{area}=2.4\pm1.1$ percent, $1.7\sigma$) and rise up again to $p_\mathrm{area}=5.3\pm0.9$ percent (panels c, $2.8\sigma$) and $p_\mathrm{area}=4.7\pm0.9$ percent (panel d, $3.6\sigma$). Similar behavior is observed throughout the whole interval. Toward the end of the event, the pulse profile first brings back its double-peaked shape (Fig.~\ref{fig:afteriX}), and only after that, with a few tens of seconds delay, the pulsed fraction drops to its normal level of $\lesssim 1$ percent, adding one more piece to the puzzle.

\begin{figure}
\begin{minipage}[h]{1.\linewidth}
\center{\includegraphics[width=1.0\linewidth,clip]{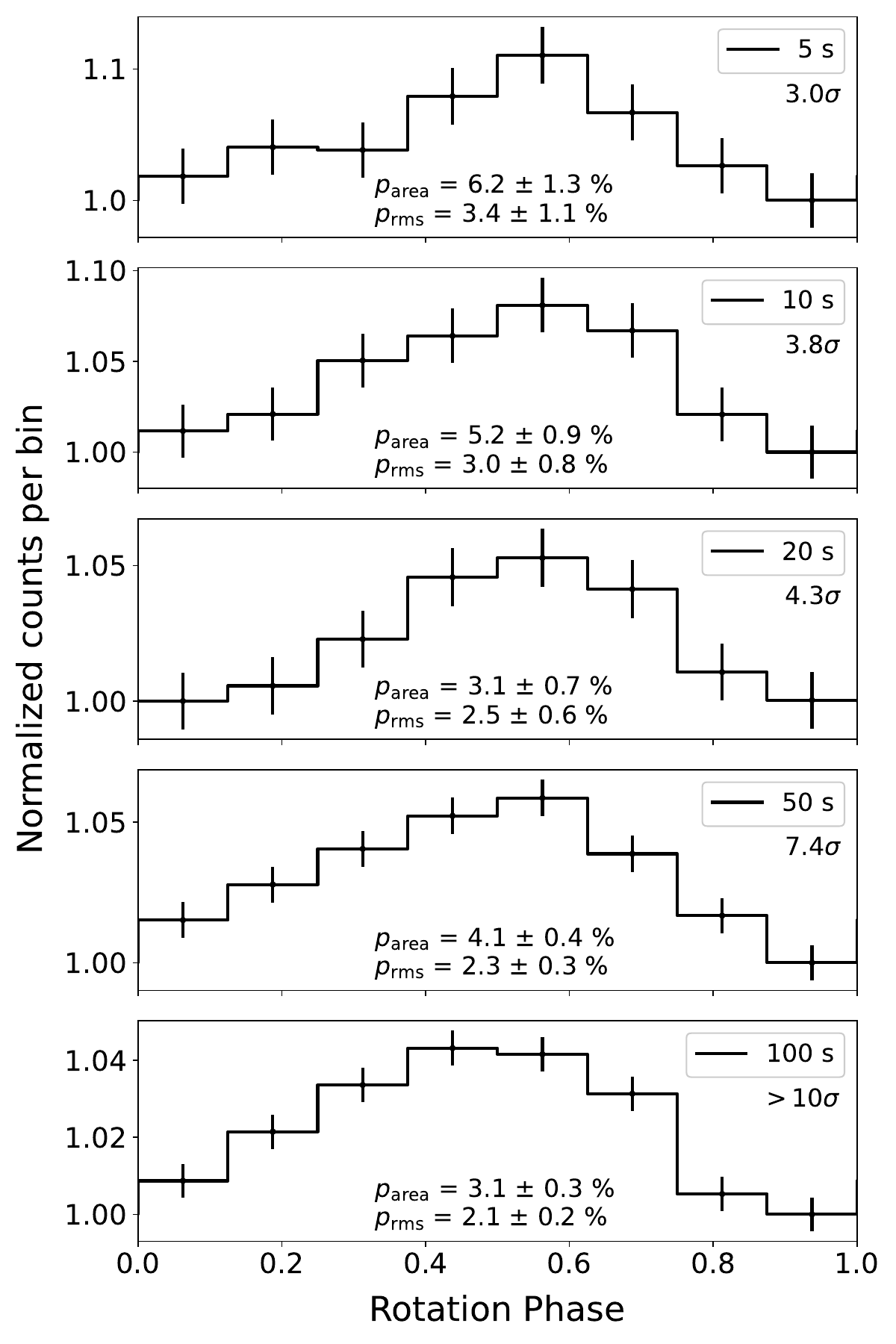}}
\end{minipage}
\caption{Top to bottom: Pulse profiles (red and blue channels combined) at different integration times (indicated in the legends) from the onset of the single-peaked interval. Corresponding background-corrected pulsed fractions and significance levels are given in each panel.}
\label{fig:iXstart}
\end{figure}

The onset of the event takes place very quickly and the pulsed fraction is the highest during the first seconds of the interval. As we can see from Fig.~\ref{fig:iXstart}, the single-peaked profile becomes prominent at the integration time as short as 5 seconds from the onset of the event and in this interval the pulsed fraction $p_\mathrm{area}$ exceeds 5 percent.

Simultaneous observations in the red and blue channels allow us to roughly analyze the behavior of the pulsar in two optical bands. The averaged background-corrected pulsed fractions of the double-peaked pulse profile are $p_\mathrm{area} = 0.69 \pm 0.05$ percent and $p_\mathrm{rms} = 0.44 \pm 0.04$ percent in the red channel, and $p_\mathrm{area} = 0.60 \pm 0.07$ percent and $p_\mathrm{rms} = 0.39 \pm 0.05$ percent in the blue channel. For the single-peaked interval, the averaged values are $p_\mathrm{area} = 3.1 \pm 0.3$ percent and $p_\mathrm{rms} = 2.1 \pm 0.2$ percent in the red channel, and $p_\mathrm{area} = 3.1 \pm 0.3$ percent and $p_\mathrm{rms} = 1.6 \pm 0.2$ percent in the blue channel, respectively.

In the absence of simultaneous spectroscopic observations, we are not able to reliably translate the observed pulsed fractions to the fluxes of the pulsed component. However, we can roughly analyze the relative evolution of the energy distribution in the pulsed emission during the observed transformation to the single-peaked state (as we show below), assessing the long-term (years) and short-term (minutes) variability of the spectrum shape. There are a number of spectra published in the literature taken after the transition of the system to the subluminous disk state. Here, we considered three of them, which we refer to as spectrum 1 \citep{Bogdanov+2015}, spectrum 2 \citep{Hernandez2016}, and spectrum 3 \citep[][with reduced spectra available\footnote{\url{https://cdsarc.cds.unistra.fr/viz-bin/cat/J/A+A/690/A344}} online]{Messa+2024}. We assumed a simple power-law energy distribution of the pulsed component, $F_\nu \sim \nu^{-\alpha}$, and first examined whether its slope varies significantly from one spectrum to another. For this purpose, for each spectrum, we calculate the flux densities at the effective frequencies of the red ($\nu_\mathrm{eff}=5.36\times10^{14}$~Hz) and blue ($\nu_\mathrm{eff}=7.01\times10^{14}$~Hz) channels. The resulting values are $F_\mathrm{eff} \simeq 5.1\times10^{-27}$~erg cm$^{-2}$ s$^{-1}$ Hz$^{-1}$ (red) and $F_\mathrm{eff} \simeq 3.8\times10^{-27}$~erg cm$^{-2}$ s$^{-1}$ Hz$^{-1}$ (blue) for spectrum 1, $F_\mathrm{eff} \simeq 8.0\times10^{-27}$~erg cm$^{-2}$ s$^{-1}$ Hz$^{-1}$ (red) and $F_\mathrm{eff} \simeq 6.1\times10^{-27}$~erg cm$^{-2}$ s$^{-1}$ Hz$^{-1}$ (blue) for spectrum 2, and $F_\mathrm{eff} \simeq 5.0\times10^{-27}$~erg cm$^{-2}$ s$^{-1}$ Hz$^{-1}$ (red) and $F_\mathrm{eff} \simeq 4.0\times10^{-27}$~erg cm$^{-2}$ s$^{-1}$ Hz$^{-1}$ (blue) for spectrum 3. Then we multiply these values by the corresponding pulsed fractions, $p_\mathrm{area}$, measured in the double-peaked and single-peaked states to obtain the flux densities of the pulsed emission, and calculate the corresponding spectral indices $\alpha$. For the double-peaked state, we find $\alpha = 1.64 \pm 0.52$ for spectrum 1, $\alpha = 1.52 \pm 0.52$ for spectrum 2, and $\alpha = 1.38 \pm 0.52$ for spectrum 3. These values are in a fairly good agreement with each other. Also, inspection of spectra obtained by \citet{Messa+2024} in high-temporal-resolution spectroscopic observations shows that the shape of the continuum does not change significantly on timescales of a few minutes. For the single-peaked state, which lasted only $\sim 3.5$ minutes, $\alpha = 1.12 \pm 0.50$ for spectrum 1, $\alpha = 1.00 \pm 0.49$ for spectrum 2, and $\alpha = 0.86 \pm 0.50$ for spectrum 3. So, the results of our analysis of the relative evolution of the spectrum during the transformation to the single-peaked state hint at hardening of the energy distribution in the pulsed component, although the uncertainties of the estimated values of $\alpha$ are large. Simultaneous spectral and timing observations of similar events are needed in the future to confirm this result.

\section{Discussion}
\label{sec:discussion}

In this paper, we report the discovery of a transformation of the optical pulse profile of J1023 from its normal, double-peaked shape to the single-peaked and back. This 220-second episode was accompanied by flaring activity observed in the light curve. In the absence of simultaneous X-ray observations, we were not able to identify the X-ray mode of the system during the event. At the same time, coordinated X-ray and optical observations show that prominent optical flares usually occur during X-ray flaring mode \citep{Papitto+2019}. Interestingly, in this paper, the authors report the detection of double-peaked pulsations during a strong 3-ks flare with the rms amplitude of $0.16\pm0.02$~percent, which is more than five times lower than during the high mode. They note that such a low value cannot be explained only by addition of an unpulsed flux to the high mode level, which means that the pulsed flux was lower during the flare. In contrast, in our event we observed a $\sim 5$ times increase of the pulsed fraction, which hints at a different nature of this particular flare. Other flares observed in our data are much shorter than the one reported in \citet{Papitto+2019}. They probably also contain pulsations with reduced amplitudes, but we were not able to distinguish them due to insufficient count statistics.

\subsection{Double-peaked pulsations}
\label{subsec:origin}

It was suggested by \citet{Papitto+2019} and \citet{Veledina+2019} that generation of the non-thermal optical and X-ray emission during the high mode is powered by collision of the pulsar wind with the accretion disk beyond the light cylinder resulting in formation of a shock, acceleration of electrons, and, consecutively, synchrotron emission in a wide range of frequencies due to motion of accelerated electrons in magnetic fields. \citet{Veledina+2019} further linked the flaring mode to rapid inflation of the inner parts of the accretion disk, which naturally explains the reduction of the pulsed fraction. This scenario appears promising, and important constraints were obtained in a number of observational works \citep[][see, however, \citealt{Jadoliya+2026}]{Campana+2019,Burtovoi+2020,Illiano+2023,Baglio+2023,Baglio+2025apj,Baglio+2025aap}. Still, many details remain to be understood. In particular, it is unclear what mechanism exactly forces the resulting emission to pulsate. In the \citet{Papitto+2019} interpretation, two synchrotron-emitting spots travel along the inner disk surface as the NS rotates, and while the emission from one half of the shock (that is closer to an observer) is absorbed by outer parts of the disk, the emission from the other half is sinusoidally modulated with the period of $P_\mathrm{s}/2$. On the other hand, \citet{Veledina+2019} noted that due to transparency of the emitting region the observer basically sees the same synchrotron source all the time, regardless of the rotation phase. They propose instead that the observed modulation is caused by the anisotropic angular distribution of the radiation emerging as a consequence of the presence of the regular large-scale magnetic field.

The observed properties of the pulsed emission allowed to estimate physical conditions inside the volume \citep[e.g.][]{Papitto+2019, Veledina+2019}. Here, we briefly reproduce these estimates since they are important for discussion of the transformation of the pulse profile to the single-peaked state. First of all, the injected energy should be radiated sufficiently fast. The characteristic cooling time scale of accelerated electrons \citep[e.g.][]{Papitto+2019} is expressed as
\begin{equation} \label{t_sync}
    t_{\rm sync} = \frac{\gamma m_{\rm e} c^2}{W_{\rm sync}} = \frac{3m_{\rm e}^3 c^5}{2e^4 B^2 \gamma},
\end{equation} 
where \citep{Ginzburg+1965}
\begin{equation} \label{synchrotron power}
    W_{\rm sync} = \frac{2 e^4 B^2}{3 m_{\rm e}^2 c^3}\, \gamma^2
\end{equation}
is the synchrotron emission power. Here, $c$ is the speed of light, $e$ and $m_{\rm e}$ are the electron charge and mass, $B$ is the magnetic field strength, and $\gamma$ is the Lorentz factor. As long as $t_{\rm sync}$ is less than roughly a quarter of the pulsar period, 
\begin{equation} \label{t_sync_cond}
    t_{\rm sync} \lesssim \frac{P_{\rm s}}{4},
\end{equation}
double-peaked pulsations are observed (provided that the size of the emitting region is small compared to $c P_{\rm s}$). On the other hand, the peak frequency of radiation from an electron with the Lorentz factor $\gamma$ is \citep{Ginzburg+1965}
\begin{equation} \label{peak freq}
    \nu(\gamma) = 0.29 \frac{3 e B \gamma^2}{4 \pi m_{\rm e} c}.
\end{equation}
These two conditions on the cooling time scale (Eq.~\ref{t_sync_cond}) and the emission frequency (Eq.~\ref{peak freq}) lead to the following expressions for $\gamma$ and $B$, 
\begin{equation} \label{gamma and B}
    \left\{
    \begin{aligned}
        & \gamma \lesssim \gamma_\mathrm{max} \simeq 3.3 \frac{e^{2/3} P_{\rm s}^{1/3} \nu^{2/3}}{m_{\rm e}^{1/3} c}, \\
        & B \gtrsim B_\mathrm{min} = B(\gamma_\mathrm{max}) \simeq 14 \frac{m_{\rm e} c \nu}{e} \frac{1}{\gamma_\mathrm{max}^2} > 1.4 \frac{m_{\rm e}^{5/3} c^3}{e^{7/3} P_{\rm s}^{2/3} \nu^{1/3}}.
    \end{aligned}
    \right.
\end{equation}
Solving this system for the period of J1023, $P_{\rm s} = 1.69$ ms, and an optical frequency of $\nu = 6 \times 10^{14}$~Hz yields $\gamma_\mathrm{max} \simeq 60$ and $B_\mathrm{min} \simeq 10^5$~G. The latter is consistent with the magnetic field strength close to the light cylinder, $B = B_{\rm s} (R_{\rm ns} / R_{\rm lc})^3$, where $B_{\rm s} \simeq 10^8$~G is the surface magnetic field of the NS, $R_{\rm ns} \simeq 10$~km is its radius, and $R_{\rm lc} = c P_{\rm s} / 2 \pi \simeq 80$~km is the radius of the light cylinder of J1023. 

The single optical electron luminosity is $L \sim 0.1$~erg/s (as follows from Eq.~\ref{synchrotron power} with $\gamma = \gamma_\mathrm{max}$ and $B = B_\mathrm{min}$) and the total optical pulsed component luminosity $L_{\rm p} \sim 10^{31}$~erg/s \citep[e.g.][]{Papitto+2019}; thus, the number of emitting particles $N_{\rm e} \sim 10^{32}$. Since the size of the emitting region, $l$, has to be smaller than $c \, P_\mathrm{s}/4 \sim 10^7$~cm for pulsations to be observed, the number density is greater than $10^{11} \text{ cm}^{-3}$. 

Pulsed emission is observed from optical up to $\sim 45$~keV X-rays \citep{Papitto+2019}, which translates to Lorentz factors of emitting electrons from $\sim 60$ to $\sim 8000,$ assuming the magnetic field strength of $\sim 10^5$~G. This energy is deposited to the inner parts of the accretion flow by the pulsar wind. The spin-down luminosity of J1023 in the radio pulsar state was $4.3 \times 10^{34}$~erg s$^{-1}$ \citep{Deller+2012}. Since, most of the time, the disk is located outside the light cylinder, we can expect that this value would not change significantly with the transition of the system to the subluminous disk state. \citet{Jaodand+2016} used X-ray observations obtained with \textit{XMM-Newton} in 2013--2015 to find that the spin-down rate of J1023 increased by $26.8(4)$ percent after the transition to the subluminous disk state. Moreover, basing on optical observations of the system obtained with \textit{Aqueye+} in 2018--2020, \citet{Burtovoi+2020} reported that the spin-down rate is only $\sim 5$ percent faster than that in the radio pulsar state. These small changes can be attributed to quite rare episodes of interaction between the accretion flow and closed magnetosphere, when the disk makes its way beneath the light cylinder. \citet{Veledina+2019} showed that even in the case of an isotropic wind, the observed high-mode X-ray luminosity of $\sim 3\times10^{33}$~erg s$^{-1}$ can be powered by the pulsar rotation energy losses under the reasonable assumption that the half-thickness of the inner edge of the disk is about 0.1 of its radius. The pulsed X-ray emission is $\lesssim 10^{32}$~erg s$^{-1}$, while the pulsed optical emission is an order of magnitude lower. At a reasonable radius of the inner edge of the disk of $2 R_\mathrm{lc}$ \citep{Papitto+2019}, the emission region of a linear size of $10^7$~cm constitutes about 0.1 of the full circumference, so the spin-down power is enough also to explain the pulsed luminosity coming from a relatively compact volume.

\subsection{Transparency conditions}
\label{subsec:transparency}
Some information, in principle, can be obtained from the spectral index $\alpha$ of the synchrotron emission. If the emission is isotropic, then the electron spectral emissivity per unit volume $J(\nu)$, on the one hand, is
\begin{equation} \label{spectral emis macro}
    J(\nu) = \frac{4 \pi D^2 F_{\nu}}{V}, 
\end{equation}
where $V \approx l^3$ is the volume of the emission region, $D$ is the distance, and $F_{\nu}$ is the pulsed flux density. On the other hand, assuming the electron energy distribution of the form $n(E)dE=\kappa E^{-p}dE$, we can write $J(\nu)$ in cgs units  \citep{Rybicki+1979}
as\begin{equation} \label{spectral emis micro}
    J(\nu) = \frac{\sqrt{3\pi}e^3 B \kappa}{2 c^2 m_{\rm e} (p+1)} \left(\frac{2\pi \nu m_{\rm e}^3 c^5}{3e B}\right)^{-(p-1)/2} a(p),
\end{equation}
where $a(p)$ is a combination of gamma functions on the order of unity. Then, $\kappa$ can be expressed from Eqs.~\ref{spectral emis macro} and \ref{spectral emis micro}, while the spectral index $p$ is related to $\alpha$ as $p = 2\alpha+1$.

\begin{figure}
\begin{minipage}[h]{1.\linewidth}
\center{\includegraphics[width=1.0\linewidth,clip]{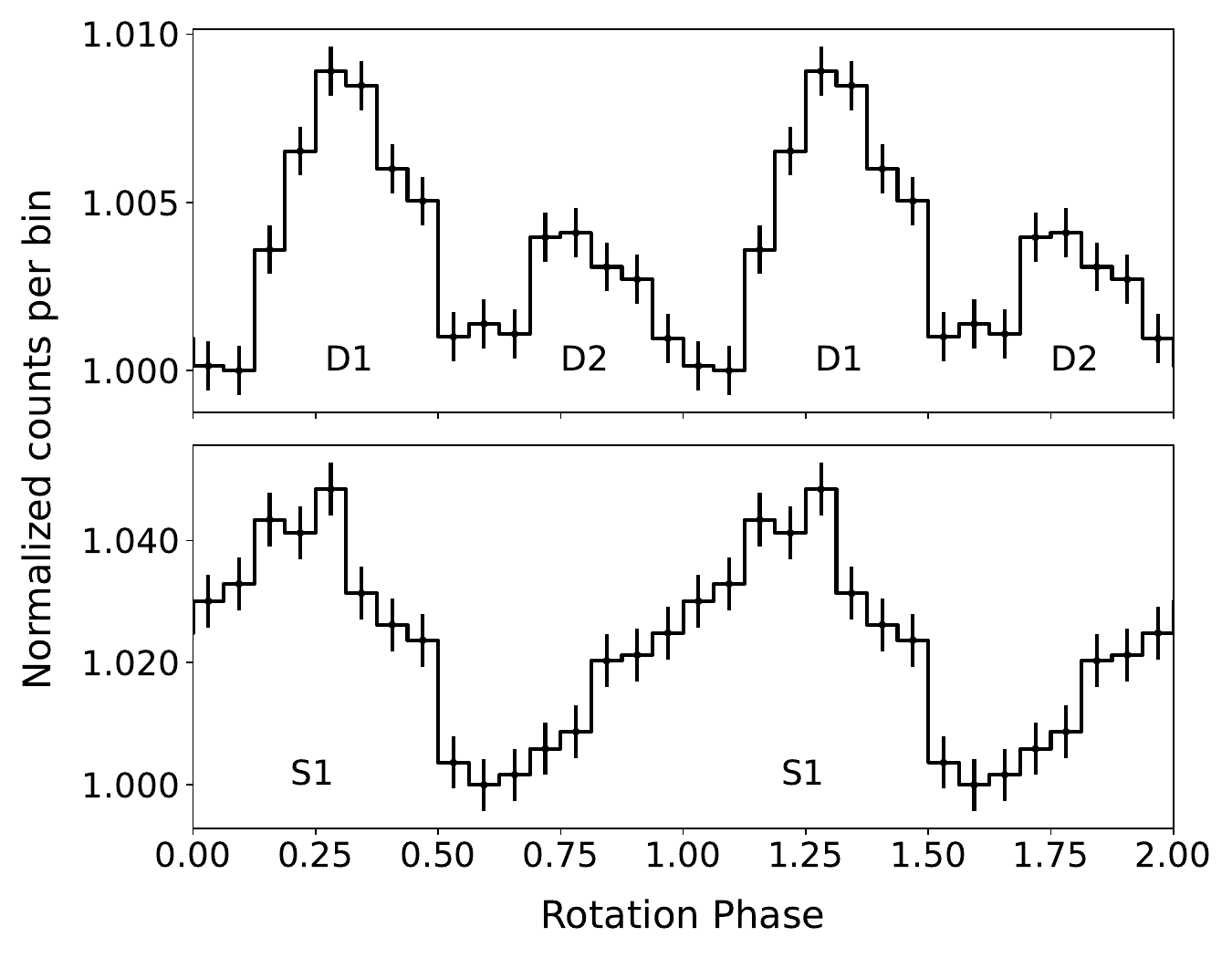}}
\end{minipage}
\caption{Top: Pulse profile (red and blue channels combined) integrated over both nights with exclusion of the single-peaked interval. ``D1'' and ``D2'' labels denote the higher and the lower peaks of the double-peaked profile. Bottom: Average pulse profile of the 220-second-long single-peaked interval. The peak is referred as ``S1''. Two phase cycles are plotted for clarity.}
\label{fig:profiles_white}
\end{figure}

If the density of the medium exceeds some threshold, it becomes opaque to the synchrotron radiation resulting in a strong intensity decrease. This effect may be related to the observed strong variability of the amplitudes of the peaks, with occasional episodes of complete suppression of pulsations. The condition of transparency is expressed as $\chi_{\nu} l \ll 1$, where $\chi_{\nu}$ is the absorption coefficient dependent on the radiation frequency \citep{Rybicki+1979}, expressed as
\begin{equation} \label{absorption coef}
    \chi_{\nu} = \frac{\sqrt{3\pi}e^3 B \kappa}{16\pi m_{\rm e}} \left( \frac{3 e B}{2\pi m_{\rm e}^3 c^5}\right)^{p/2} \nu^{-(p+4)/2} \; b(p),
\end{equation}
where $b(p)$ is another combination of gamma functions on the order of unity.  Substituting $\kappa$ yields
\begin{equation}
    \chi_{\nu} = \left(\frac{3B e}{128c m_{\rm e}^3 \nu^5}\right)^{1/2} \frac{4 \pi D^2 F_{\nu}}{V} \,(p+1)\, \frac{b(p)}{a(p)},
\end{equation}
and for $p \approx 2.4$, $b(p=2.4)/a(p=2.4) \approx 1$, we have
\begin{equation}
    \chi_{\nu} l\approx 0.02 \times \left(\frac{B}{10^5 \text{ G}}\right)^{1/2} \nu_{\rm red}^{-5/2}\, D^2_{1.37}\, F_{5}\, l_7^{-2}\, p_{2.4},
\end{equation}
where $D_{1.37}=D/(1.37\text{ kpc})$ \citep{Deller+2012}, $\nu_{\rm red} = \nu/(5.3\times10^{14}\text{ Hz})$ is the effective frequency of the red detector, $F_5 = F_\nu/(5\mu\text{Jy})$ is the pulsed flux density in the red channel estimated from the optical spectrum of J1023 \citep{Hernandez2016} and the pulsed fraction $p_\mathrm{area}$.
Let us rewrite this condition in terms of electron number density. Spectral emissivity per unit volume given by Eq.~\ref{spectral emis micro} can also be written as 
\begin{equation}
    J(\nu) \approx W_{\nu} \frac{N_{\rm e}}{l^3}, 
\end{equation}
where in the frequency range of interest, $W_{\nu} \approx W_{\rm sync} / \nu$. Then, 
\begin{equation}
    \chi_{\nu} l \approx 0.01 \times \left(\frac{B}{10^5 \text{ G}}\right)^{5/2} \left(\frac{N_{\rm e}}{10^{32}}\right) \left(\frac{\gamma}{60}\right)^{2} \,\nu_{\rm red}^{-7/2} \,l_7^{-2}.
\end{equation}
While the obtained value is rather small, it strongly depends on the characteristics of the emitting volume (in particular, its size $l$). Thus, in summary, the upper limit on $l$ is set by the pulsar period, while the lower limit follows from the transparency condition, roughly constraining $l$ in the range of $10-100$~km (for specified $B$, $\gamma$, and $N_\mathrm{e}$).

\subsection{Transformation scenarios}
\label{subsec:transformation}
Visually inspecting the behavior of the double-peaked and the single-peaked pulse profiles with respect to the rotation phase (see Fig.~\ref{fig:profiles_white}), we note that (1) the phase of the minimum of the single-peaked profile is approximately the same (roughly in the range of 0.55--0.65) as the phase of the minimum between D1 and D2 peaks of the double-peaked profile; and (2) the descending edge of the single-peaked profile is similar to the descending edge of D1 peak of the double-peaked profile (both cover the range of phases of about 0.30--0.55). The shape of the peaks is the result of complex interplay between different factors, such as the time of propagation of the injected energy inside the emitting region, the synchrotron cooling time $t_\mathrm{sync}$, the geometric configuration, and so on. Changes in the physical conditions may affect the shape of D2 peak, stretching its leading and trailing edges and resulting in the observed quick merging of D1 and D2 peaks into S1. From Eqs.~\ref{t_sync}--\ref{gamma and B}, it follows that slight changes in $\gamma$ and $B$ strongly affect the cooling time scale. We should note, however, that both pulses are generated by roughly the same volume when it is hit by diametrically opposed ``lobes'' of the pulsar wind, and why the shape of D2 peak is affected by these changes while the shape of D1 is not remains unclear.

\begin{table}
\renewcommand{\arraystretch}{1.2}
\caption{Parameters of the emitting plasma for D1 and D2 peaks in the single-peaked and double-peaked states within the toy model described in the text.}
\label{tab:peaks}
\begin{center}
\begin{tabular}{ccccc}
\hline
Parameter & \multicolumn{2}{c}{Double-peaked} & \multicolumn{2}{c}{Single-peaked} \\
& D1 & D2 & D1 & D2 \\
\hline
$\gamma_\mathrm{max}$ & \multicolumn{2}{c}{59} & 59 & 74 \\
$B_\mathrm{min}$ (kG) & \multicolumn{2}{c}{140} & 140 & 88 \\
$N_\mathrm{e}$ ($10^{31}$) & 5.4 & 3.6 & 22 & 36 \\
$n$ ($10^{10}$ cm$^{-3}$) & \multicolumn{2}{c}{$\sim 4.5$} & \multicolumn{2}{c}{$\sim 30$} \\
\hline
\end{tabular}
\tablefoot{The number densities, $n$, are calculated for $V = l^3 = 10^{21}$~cm$^{3}$ and the values averaged between the two peaks are given.}
\end{center}
\end{table}

Let us consider the changes of the parameters of the emitting plasma within a toy model, assuming that the transformation to the single-peaked state is only governed by the broadening of D2 peak and not by a complete reconfiguration of the structure of the pulsar wind-disk shock. We suppose the total optical pulsed luminosity of $\sim 10^{31}$~erg s$^{-1}$ is apportioned between D1 and D2 as $6 \times 10^{30}$~erg s$^{-1}$ and $4 \times 10^{30}$~erg s$^{-1}$ in the double-peaked state (roughly according to the relative areas under the peaks), while the luminosity of $\sim 4\times 10^{31}$~erg s$^{-1}$ (since the pulsed fraction is about 4 times higher) is equally apportioned between D1 and D2 ($2 \times 10^{31}$~erg s$^{-1}$ each) in the single-peaked state. These values are somewhat arbitrary due to strong variability of the peaks, but here we are interested in general behavior of $\gamma$, $B$, and $n$ with respect to the observed properties of pulsations. Additionally, we suppose that the trailing edges of the peaks are governed by the synchrotron cooling timescale, $t_\mathrm{sync}$. Then, using Eqs.~\ref{synchrotron power} and \ref{gamma and B}, we can write $\gamma_\mathrm{max} = 786 \,t_\mathrm{sync}^{1/3}$, $B_\mathrm{min} = 787 \,t_\mathrm{sync}^{-2/3}$~G and $L_\mathrm{p} = 6.1 \times 10^{-4} N_\mathrm{e} t_\mathrm{sync}^{-2/3}$~erg s$^{-1}$, where $t_\mathrm{sync}$ is given in seconds. In the double-peaked state, $t_\mathrm{sync} \simeq P_\mathrm{s} / 4 = 0.42$~ms for both D1 and D2 peaks, while in the single-peaked state it stays the same for D1 and roughly doubles for D2. The corresponding parameters of the emitting plasma for the peaks in these two states are listed in Table~\ref{tab:peaks}. Our toy model thus suggests that if in the double-peaked state the values of $\gamma$ and $B$ are close to $\gamma_\mathrm{max}$ and $B_\mathrm{min}$, then moderate variations of these parameters by about 30 percent may lead to broadening of D2 peak. At the same time, these variations does not significantly change the luminosity $L_\mathrm{p}$, which mainly depends on the number of the emitting particles, $N_\mathrm{e}$. The increase by a factor of $6-7$ in the number densities may be caused, for instance, via the accumulation of matter in the inner parts of the disk due to temporal increase in the accretion rate, while moderate differences in the values of $N_\mathrm{e}$ for D1 and D2 peaks may be related to slightly different sizes of the shocks generated by the opposed lobes of the pulsar wind. \citet{Veledina+2019} associate flares with a rapid increase of the disk thickness at the inner edge. In this case a larger fraction of the pulsar wind is intercepted by the disk, which may lead, among other things, to changes of physical conditions inside the emitting volume. However, the pulsed fraction is expected to drop in such events \citep[as was observed during a strong 3-ks flare in ][]{Papitto+2019}. The mechanism for a five-time increase of the pulsed fraction during the single-peaked interval still remains to be understood.

Transformation of the pulse profile in principle could be related to global reconfiguration of the magnetosphere of the NS. Such events have been discussed in the context of magnetar outbursts \citep[e.g.,][]{Eichler2002}. A complex pulse profile behavior was observed in magnetar flares and afterglows \citep{Mazets+1999,Feroci+2001,Palmer+2005}. Recently, peculiar mode changes and pulse profile phase shifts of the $\gamma$-ray pulsar PSR J2021+4026 were associated with reconfigurations of the dipole and quadrupole components of the magnetic field \citep{Razzano+2023,Fiori+2024}. However, it is difficult to conceive that something similar could happen in a very old NS with a much lower magnetic field on the order of $10^8$~G. Some kind of interaction between the charged particles produced in the pulsar magnetosphere and the portion of the infalling plasma that managed to reach the closed magnetosphere and penetrate into it probably could change the morphology of the pulsar wind without the reconfiguration of the magnetosphere itself. However, this scenario remains rather unclear and requires further investigation.

\section{Summary}
\label{sec:sum} 

We report the discovery of the rapid transformation of the optical pulse profile of the transitional MSP J1023 from its normally observed double-peaked shape to the single-peaked and back again, which we were able to precisely trace in our high temporal resolution observations. The transformation occurred on a timescale of a few seconds or less. At the start of the single-peaked interval, the pulsed fraction exceeded 5 percent, which is almost an order of magnitude larger than the average pulsed fraction in the normal state with the double-peaked pulse profile. Toward the end of this peculiar 220-second-long event, the pulse profile started to take its normal shape and it was only afterward (with a few tens of seconds delay) that the pulsed fraction dropped back to the level of $\lesssim 1$ percent. Strong flaring activity was observed during this event. Unfortunately, the lack of simultaneous X-ray observations makes it very difficult to understand the processes going on in the system during this change in state.

The shock near the light cylinder of the NS, where the pulsar wind hits the infalling matter, is usually considered as the source of the pulsed emission. The interaction of diametrically opposed lobes of the pulsar wind with the inner radius of the accretion disk gives rise to two emission peaks per one rotation of the NS. At the same time, the details of this scenario remain rather unclear. An aspect that is even more puzzling is the temporary transformation of the pulse profile presented in this paper. It might be related to changes of the physical conditions inside the emitting volume resulting in merging of the two peaks into one. More exotic scenarios include global reconfiguration of the NS magnetosphere or pattern change of the pulsar wind.

Our study reveals one more enigmatic feature of J1023. We stress the importance of further monitoring of the system, especially with regard to the need to carry out coordinated X-ray and optical timing and spectral observations that will make it possible to perform more detailed analyses of similar events in the future.

\section*{Data availability}

Our optical data are available at the CDS via anonymous ftp to \url{cdsarc.u-strasbg.fr} (130.79.128.5) or via \url{http://cdsweb.u-strasbg.fr/cgi-bin/qcat?J/A+A/}. The \textit{Swift} data are available via \url{https://www.swift.ac.uk/}.

\begin{acknowledgements}
We dedicate this publication to the memory of our dear colleague, Yuri Shibanov, who passed away on September 14, 2025. His contributions to pulsar research in various wavelength ranges, both theoretical and observational, and in particular to the study of PSR J1023+0038, are exceptionally important in modern astrophysics. He also possessed remarkable human qualities~-- openness, kindness, willingness to share his knowledge and to support his students and colleagues.

We are grateful to the anonymous referee for important remarks that helped to significantly improve the quality of the paper. SVK was supported by the EU and by the Czech Ministry of Education, Youth and Sports (project CZ.02.01.01/00/22\_008/0004632 -- FORTE). The work of AST was supported by the Russian Science Foundation Grant No. 24-12-00320. The work of GMB and VLP was carried out within the framework of the state assignment of the Special Astrophysical Observatory of the Russian Academy of Sciences, approved by the Ministry of Science and Higher Education of the Russian Federation. DAZ was supported by the baseline project FFUG-2024-0002 of the Ioffe Institute.
This study is based on the data obtained at the unique scientific facility the Big Telescope Alt-azimuthal of SAO RAS equipped with the MANIA complex. This work made use of data supplied by the UK Swift Science Data Centre at the University of Leicester.
DAZ thanks Pirinem School of Theoretical Physics for hospitality.
\end{acknowledgements}

\bibliographystyle{aa}
\bibliography{ref}

\appendix

\section{\textit{Swift/XRT} observations}
\label{a:swift}

Properties of X-ray emission of J1023 is one of the best probes of its current state. In the absence of simultaneous X-ray observations we are not able to compare the behavior of the system in the optical and X-ray bands for our data set and, in particular, during the puzzling episode of the structural change of the pulse profile. However, it is worth checking the state of the system near our epoch.

\begin{figure}
\begin{minipage}[h]{1.\linewidth}
\center{\includegraphics[width=1.0\linewidth,clip]{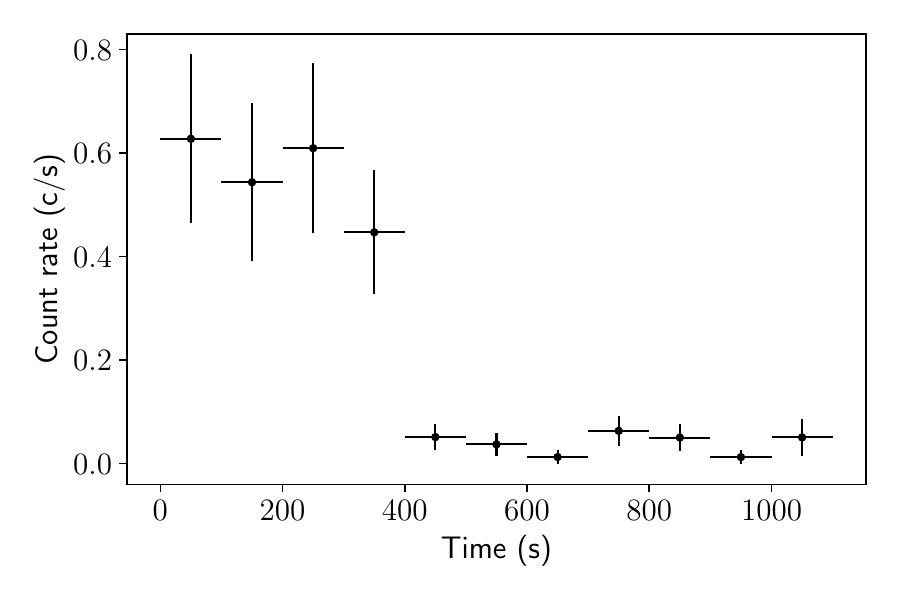}}
\end{minipage}
\caption{\textit{Swift/XRT} light curve in the 0.3--10~keV range. Zero time corresponds to 28 Nov 2017 01:16:06 UT, two weeks after our observations. The time bins are 100 seconds wide. Apparently, a switch from the high to the low mode occurred at around 400 seconds from the start.}
\label{fig:Xray}
\end{figure}

\textit{Swift/XRT} observed J1023 in the Photon Counting mode on November 28 (ObsID 00033012183), two weeks after our observations. The count rate during the 1.1 ks exposure was $0.131 \pm 0.013$~s$^{-1}$ in the 0.3--10.0~keV energy range. Inspection of the light curve (Fig.~\ref{fig:Xray}) hints that after first $\sim$ 400 seconds the system probably switched from the high to the low mode. 

We extracted the spectrum in the 0.3--10~keV range using the \textit{Swift/XRT} data products generator \citep{Evans+2009}, and fitted it with an absorbed power law model using the X-Ray Spectral Fitting Package ({\sc xspec}) v.12.11.1 \citep{arnaud1996}. The hydrogen column density along the line of sight cannot be reliably constrained from this observation due to the low count statistic. We fixed it at the value of $N_\mathrm{H} = 3.1 \times 10^{20}$~cm$^{-2}$ reported by \citet{Bogdanov+2015} for the high mode. We obtained a good fit with the spectral photon index $\Gamma = 1.8 \pm 0.2$ and the unabsorbed flux $F_\mathrm{X} = (4.6 \pm 0.8) \times 10^{-12}$ erg cm$^{-2}$ s$^{-1}$ in the 0.3--10~keV range, which at the parallax distance of 1.37 kpc \citep{Deller+2012} translates to the luminosity $L_\mathrm{X} = (1.0 \pm 0.2) \times 10^{33}$ erg s$^{-1}$. We conclude that the system was in its normal subluminous disk state during the \textit{Swift/XRT} observation performed two weeks after our optical observation.

\end{document}